\documentclass[12pt,revtex]{article}
\begin{document}
\begin{center}
{\bf\Large 
Electric multipole plasmons in deformed sodium clusters
}

\vspace{1cm}
W. Kleinig (1 and 2), V. O. Nesterenko (1), and P.-G. Reinhard (3)

\vspace{1cm}
(1) Bogoliubov Laboratory of Theoretical Physics,
Joint Institute for Nuclear Research, Dubna, Moscow region,
141980, Russia

(2) Technische Universit\"at Dresden,
Institut f\"ur Analysis, Dresden, D-01062, Germany

(3) Institut f\"ur Theoretische Physik,
Universit\"at Erlangen, D-91058, Erlangen, Germany

\vspace{1cm}
 ABSTRACT
\end{center}

\vspace{1cm}
 The random-phase-approximation (RPA) method with {\it
separable} residual forces (SRPA) is proposed for the description
of multipole electric oscillations of valence electrons in {\it
deformed} alkali metal clusters. Both the deformed mean field and
residual interaction are derived self-consistently from the
Kohn-Sham functional. SRPA drastically  simplifies the
computational effort which is urgent if not decisive for deformed
systems.
The method is applied to the description of dipole, quadrupole and
octupole plasmons in deformed sodium clusters of a moderate size.
We demonstrate that, in clusters with the size $N>50$,
Landau damping successfully competes with deformation splitting
and even becomes decisive in forming the width and gross
structure of the dipole plasmon. Besides, the plasmon is
generated by excitations from both ground state and shape isomers.
In such clusters familiar experimental estimates for deformation
splitting  of dipole plasmon are useless.

\vspace{1cm}

\newpage
\section{Introduction}

Most of free atomic clusters are known to exhibit quadrupole,
hexadecapole, octupole, etc. deformations (see, e.g.,
\cite{Clem_def}-\cite{Bon_def} and references therein). The
deformations persist from lightest \cite{Man_def,Bon_def} to
heavy\cite{Pash_def} clusters and considerably influence cluster's
static and dynamical properties.  For example, the axial (triaxial)
quadrupole deformation leads to the splitting of the dipole plasmon
into two (three) peaks \cite{Sel_def,Ell_na11,Sch_def} and to specific
low-energy orbital magnetic dipole
resonance \cite{Lip_PRL}-\cite{Ne_Rich}.

Investigation of deformation effects in the dynamics of clusters
is a topic of current interest, for reviews and monographs see
\cite{Ne_SN}-\cite{ownrev}.
At the same time, the handling of
excitations in deformed clusters is much more tedious because there
are less selection rules and thus larger irreducible particle-hole
spaces.  In particular, random-phase-approximation (RPA) calculations
for multipole oscillations of the valence electrons (plasmons) require
diagonalization of matrices with a very high rank
(up to $10^5$ and more).

For this reason, the calculations are mainly limited either to
light clusters (with number of atoms $N<20$) where the computational
effort is still acceptable \cite{Bon_def}, or to very heavy clusters
where one may neglect effects of quantum shells and use macroscopic
models, thus avoiding a deal with a huge configuration space. The main
computational troubles are concentrated in the intermediate size
region, $10<N<10^3$, where sizable effects of quantum shells take
place.  Here there exits only few calculations with RPA or
fully fledged Time-Dependent Local Density Approximation (TDLDA)
(see, for example, \cite{Eka}-\cite{Ull_Na92}).
Thus one often
chooses simplified methods, like Sum-Rule Approach (SRA)
(see, e.g., \cite{Se_SRA}) or local RPA (LRPA) \cite{LRPA1,LRPA2}.
They allow to reproduce the energy and gross
structure of the plasmon, but cannot properly describe such important
feature of cluster dynamics as Landau damping.
At the same time, it is just Landau damping which greatly influences
the shape of the plasmon and provides the dominant contribution to
its width in medium and heavy clusters \cite{LRPA2}-\cite{Ne_EPJ}.
So, there is the need in RPA approach for deformed clusters,
which, on the one hand, gives comprehensive and accurate description
of multipole plasmons, including Landau damping effects, and, on the
other hand, accomplishes this in economical way.

To this end, a new kind of RPA method has been recently
proposed \cite{babst,Ne_PRA}. It invokes a {\it self-consistent}
expansion of the residual two-body interaction
$\hat{V}_{\rm res}$ into a sum of separable terms:
\begin{eqnarray}  \label{eq:separ}
  \langle p_1,p_2|\hat{V}_{\rm res}|h_1,h_2\rangle
  &\Rightarrow&
  \sum_{k,k'=1}^K \kappa_{kk'}
    \langle p_1|\hat{Q}_k|h_1\rangle
    \langle p_2|\hat{Q}_{k'}|h_2\rangle
  \;.
\end{eqnarray}
\noindent
Here, $\kappa_{kk'}$ is the matrix of strength constants,
$\langle p|\hat{Q}_k|h\rangle$ is the particle-hole matrix element of
the local one-body operator $\hat{Q}_{k}(\mathbf{r})$.  The expansion
index $k$ labels various angular momentum channels $\lambda\mu$
as well as a series of different radial functions within each channel.
The method proposed is called as
a "separable RPA" or SRPA. Due to the separable ansatz, SRPA has the
advantage to transform the high-rank RPA matrix to a simple dispersion
equation, which drastically reduces the computational effort.  This is
crucial for studies of the linear response in deformed or large
spherical systems, when one has to deal with a huge particle-hole
configuration space.

  It should be emphasized that SRPA has important distinctions as
compared with trivial separable RPA schemes
widely used in many-body studies, especially in nuclear theory \cite{Ring,Sol}.
Such schemes exploit only one
separable term with an intuitive guess for the separable one-body
operator ${\hat Q}(\mathbf{r})$, and the strength constant $\kappa$ is fitted
as to reproduce available experimental data. The present SRPA
constitutes a systematic expansion of the self-consistent
residual interaction. The analytical expressions for both
strength matrix $\kappa_{kk'}$ and operators $\hat{Q}_{k}(\mathbf{r})$
are uniquely prescribed by the given $\hat{V}_{\rm
res}$. There are thus no adjustable parameters. One can easily control
the convergence of the expansion by comparing
calculations with different cutoffs $K$. In fact, already a single SRPA
separable term  provides a
fair description of the gross structures, and a good handful of
terms allows to reproduce exact RPA calculations
\cite{babst,Ne_EPJ}. The SRPA scheme was also proven to cooperate
with the more involved Hamiltonians containing non-local
pseudo-potentials \cite{Ne_EPJ}.

The separable expansions \cite{babst} and \cite{Ne_PRA} use different
self-consistent prescriptions. The method \cite{babst} deals with
simple $\hat{Q}_{k}(\mathbf{r})$ operators but for the price of more
separable terms. Vice versa, the operators in \cite{Ne_PRA} are more
involved but the method needs less number of separable terms.
In the present paper, we will deal with the method \cite{Ne_PRA}
where the form of the separable operators $\hat{Q}_k(\mathbf{r})$
and strength matrix $\kappa_{kk'}$ are determined
from the Vibrating Potential Model (VPM) \cite{Row}-\cite{Ne_ZPD}.
The number of separable terms $K$  is usually limited by 2-3 and
4-6 in light and heavy clusters, respectively.

First SRPA results for the dipole plasmon and scissors mode in
deformed clusters were presented in refs. \cite{Ne_PRL}-\cite{Ne_SN},
\cite{Ne_ZPD}-\cite{Ne_Pra}.
But these calculations were based on a simple shell-model ground state
using a deformed Woods-Saxon potential where the deformations were
not obtained within the same calculation scheme but taken from
other sources,
both experimental and theoretical. In the present paper, we present
fully self-consistent SRPA scheme for deformed
clusters. Both, the deformed mean field and residual interaction, are
derived from the same Kohn-Sham functional. Ground state deformations
are optimized by minimization of the total energy.  SRPA is applied to
detailed description of the dipole, quadrupole and octupole
plasmons in axially-deformed singly-charged $Na$ clusters of a different
size. Both quadrupole and hexadecapole deformations are taken into
account.

We will show that in light clusters
the dipole plasmon exhibits distinctive two-peak structure caused by
the deformation splitting and just this splitting is mainly
responsible for the plasmon width. The ratio between the peaks allows
to conclude on either prolate or oblate shape takes place. The
situation changes for medium and heavy deformed clusters where Landau
damping becomes the main mechanism defining the width and gross
structure of the plasmon.  Moreover, such clusters exhibit a
complicated energy surface with shape isomers which, in spite of a
tiny energy deficit relative to the ground state, possess quite
different deformation. Obviously, such isomers can contribute to the
plasmon. Due to both the factors, Landau damping and contributions of
shape isomers, gross structure of the dipole plasmon in medium and
heavy deformed clusters cannot already serve as a {\it direct}
fingerprint of the deformation and familiar experimental
prescriptions for getting
magnitude and kind (prolate or oblate) of cluster's quadrupole
deformation become useless.

In Section 2, the deformed Kohn-Sham mean field is described.
The existence of shape isomers in heavy clusters is demonstrated.
SRPA equations are derived and specified for
deformed clusters. In Sec. 3, SRPA results for the dipole
plasmon are discussed. The same is done for quadrupole
and octupole plasmons. The summary is given in Sec. 4.
Appendixes A-C contain details of the formalism.

\section{Deformed Kohn-Sham mean field}

\subsection{Basic constituents of  description}

We start with the Kohn-Sham functional for
a cluster including $N_e$ valence electrons and $N$ atoms
\begin{equation} \label{eq:func}
E_{\rm tot}(t)= E_{\rm kin}(t) + E_{xc}(t) + E_C(t)
\end{equation}
where
\begin{eqnarray}
  E_{\rm kin}(t)
  &=&
  \frac{\hbar^2}{2m_e} \int d\mathbf{r}\, \tau ({\bf r},t)
  \;,
\label{eq: E_T}\\
  E_{xc}(t)
  &=&
  \int d{\bf r}\,\rho({\bf r},t)\,
   \epsilon_{xc}(\rho\left({\bf r},t)\right)
  \;,
\label{eq:exc} \\
  E_C(t)
  &=&
  \frac{e^2}{2}\int \int  d{\bf r}d{\bf r}_1
  \frac{(\rho({\bf r},t) - \rho_i({\bf r}))
        (\rho({\bf r}_1,t) - \rho_i({\bf r}_1))}
       {\vert {\bf r} - {\bf r}_1\vert}
\label{eq:ec}
\end{eqnarray}
\noindent
are the kinetic energy,
exchange-correlation (in the local density approximation
\cite{GL}, see Appendix A for details), and Coulomb terms,
respectively. Here,
$\rho_i({\bf r})$ is the ionic density in the
jellium approximation, $\rho({\bf r},t)$ and $\tau ({\bf r},t)$
are, respectively, the density and kinetic energy density of valence
electrons, expressed through the single-particle wave functions
$\psi _{q} ({\bf r},t)$ as
\begin{eqnarray} \label{eq:edens}
  \rho({\bf r},t)
  &=&
   \sum_{q}^{occ} w_q\vert{\psi }_{q}({\bf r},t)\vert^2
  \quad,\quad
 \tau({\bf r},t)=
\sum_{q}^{occ} w_q\vert \nabla{\psi }_{q} ({\bf r},t)\vert ^2
  \quad
\end{eqnarray}
with a temperature $T$ introduced through thermal occupation
weights
\begin{equation}
 w_q=\frac{1}{1+exp(\frac{\varepsilon_q - \lambda_F}{T})}
 \;.
\end{equation}
Here, $\varepsilon_q$ is the energy of the single-particle state $q$.
The chemical potential $\lambda_F$ serves to ensure the
correct number of valence electrons in the static
ground state, i.e.  $\int d\mathbf{r}\rho_0({\bf r})=N_e$.
The temperature regulates the occupation numbers of the
single-particle states, which becomes important for clusters with
partially occupied subshells.  In medium and heavy clusters where the
energy surface is very complicated the introduction of a temperature
softens numerical fluctuations and thus improves the convergence
of numerical results.

The Kohn-Sham single-particle Hamiltonian is obtained by variation
\begin{equation} \label{eq:Hsp}
  \hat{h}({\bf r},t) \psi _{q} ({\bf r},t)
  =                                 
  \frac{\delta E_{tot}(t)}{\delta \psi^{*} _{q} ({\bf r},t)}
  \;.
\end{equation}
This yields
\begin{eqnarray}
  \hat{h}({\bf r},t)
  &=&
  -\frac{\hbar^2}{2m}\bigtriangledown^2 + U_{xc}({\bf r},t)+U_C({\bf r},t)\: ,
\label{eq:H0}\\
\label{eq:UXC}
 U_{xc}({\bf r},t)
  &=&
  \epsilon_{xc}(\rho({\bf r},t))
  +
  \rho({\bf r},t)
  \frac{\delta\epsilon_{xc}\left(\rho({\bf r},t)\right)}{\delta\rho}
  \;,
\\
\label{eq:UC}
U_C({\bf r},t)
  &=&
  e\int d{\bf r}_1\frac{\rho({\bf r}_1,t) - \rho_i({\bf r}_1)}
  {\vert {\bf r} - {\bf r}_1\vert }
  \;,
\end{eqnarray}
see Appendix A for more details.

The ionic background is described in the soft jellium approximation.
The density of the deformed jellium is
\begin{equation} \label{eq:densi}
\rho_i({\bf r})=\frac{\rho_{i0}}{1+exp((r-R(\Theta ))/\alpha )}
\end{equation}
where quadrupole and hexadecapole deformations,
$\delta_2$ and $\delta_4$,
are introduced through cluster's radius as
\begin{equation}\label{eq:R}
R(\Theta )=R_0(1+\sum_{\lambda =2,4} \delta_{\lambda}Y_{\lambda
0} (\Theta )).
\end{equation}
Here $R_0=Cr_s N^{1/3}$; $r_s$ is the Wigner-Zeitz radius
($r_s=3.96$ a.u. for Na clusters); the
coefficient $C$ is adjusted to
ensure volume conservation
$\int d{\bf r} \rho_i({\bf r})=N$,
$\rho_{i0}=\frac{3}{4\pi r_s}$ is the bulk density.
The diffuseness values $\alpha =1$ and 0.8 a.u. are chosen
for calculations with high ($T=400-800\; K$) and low ($T=105\; K$)
temperature, respectively. These values allow to describe
the energy of the dipole plasmon in a wide size
region ( see Ref. \cite{Ne_EPJ} and present results). Decrease of
the diffuseness parameter results in a redshift of the dipole plasmon
which should follow lowering the temperature.
The recent experimental data for the dipole plasmon in light and medium
clusters \cite{exp} were obtained at $T=105\; K$. So, just this
temperature and the value $\alpha =0.8$ a.u. are used in the paper for
all clusters except of the heaviest one $Na^+_{119}$. For the
latter cluster the experimental data are absent and so
the common value $\alpha =1.0$ a.u. is used.

\subsection{Representation of wave functions for stationary deformed states}

Stationary states of the cluster are obtained as
solution of the stationary Kohn-Sham equations
$\hat{h}_0({\bf r})\psi_q({\bf r})=\varepsilon_q\psi_q({\bf r})$
where $\hat{h}_0({\bf r})$ is the single-particle
Hamiltonian (\ref{eq:Hsp}) in the static limit.
We expand single-particle wave
functions of the deformed Kohn-Sham potential
in the complete basis of eigenfunctions
$\phi_{s}({\bf r})\equiv\phi_{nl\Lambda}({\bf r})$
of the same but spherical Kohn-Sham potential:
\begin{equation}
\label{eq:psi}
\psi_{q={\cal N}n_z \Lambda}({\bf r})=
\sum_{nl} a^{q}_{nl} \phi_{nl\Lambda}({\bf r}).
\end{equation}
The spherical states
$\phi_{s}({\bf r})\equiv\phi_{nl\Lambda}({\bf r})
=R_{nl}(r) Y_{l\Lambda}(\Omega)$
are characterized  by quantum numbers $s=
nl\Lambda$ (the node number, orbital moment and its projection to
the symmetry axis, respectively), where the orbital moment is
connected with the space parity as $\pi =(-1)^l$. For the sake of
simplicity, spin functions are omitted
though, of course, their contribution is taken into account
in all relevant values. Deformed states are specified by
asymptotic Nilsson-Clemenger
quantum numbers $q ={\cal N}n_z \Lambda$ \cite{Clem_def}
where $\cal N$ is the principle
shell quantum number (the total number of quants,
${\cal N}=n_x+n_y+n_z$) and $\pi =(-1)^{\cal N}$.
The levels in axially-deformed
clusters are twofold (for $\Lambda =0$) or fourfold (for $\Lambda
\ne 0$) degenerated.
The electronic density reads in this representation
\begin{equation} \label{eq:dedens}
  \rho_0({\bf r})
  =2\sum_{q={\cal N}n_z \Lambda}^{occ} (2-\delta_{\Lambda ,0}) w_q
   \vert \psi_{q}({\bf r})\vert ^2.
\end{equation}
Making use the multipole expansion for static potentials
(\ref{eq:UXC})-(\ref{eq:UC}) and densities
(\ref{eq:densi})-(\ref{eq:dedens}),
the equations to determine single-particle energies
$E_{q}$ and wave functions $\psi_{q}$ are derived.
As a first step, the Kohn-Sham problem is solved in
the spherical limit
and then the basis obtained is used for deformed Kohn-Sham potential.
At each iteration, the conservation of $N_e$ and $N$ is controlled
by fitting the parameters $\lambda_F$ and $C$. The procedure is performed for
every point in the deformation map $\{\delta_2,\delta_4\}$.
The equilibrium deformation parameters $\delta_2$ and $\delta_4$
are determined by minimization of the total energy
(\ref{eq:func}). Details of the calculation scheme are given in
the Appendix A.

\subsection{Results for stationary states of deformed clusters}

We consider axially deformed clusters $Na_{15}^+$,
$Na_{27}^+$, $Na_{35}^+$, $Na_{53}^+$, and $Na_{119}^+$. These
clusters cover different size regions and represent both prolate
and oblate shapes. Following calculations
\cite{Sel_def}-\cite{Thom_def}, they do not exhibit neither
triaxiality, nor octupole deformation and, so, we may safely deal
only with axial quadrupole and hexadecapole deformations. The
equilibrium values of $\delta_2$ and $\delta_4$ and the
corresponding moments are presented in Table I.  The moments are
defined as
\begin{eqnarray}
 \beta_{\lambda}
 &=&
 \frac{4\pi}{3}
 \frac{\int d{\bf r}\rho_0({\bf r}) r^{\lambda}Y_{\lambda 0}}{N_e {\tilde R}}
\label{eq:QL}
 \;,\quad
 {\tilde R}
 =
 \sqrt{\frac{5}{3} \frac{\int d{\bf r}\rho_0({\bf r}) r^2 }
      {\int d{\bf r}\rho_0({\bf r})}}
 \;,\quad \lambda =2,4
 \;.
\nonumber
\end{eqnarray}
As we checked, the calculated moments are close to those obtained
within the Structure Averaged Jellium Model (SAJM) \cite{MR_95}.
Table I shows that some clusters exhibit considerable hexadecapole
deformation. Further, in particular clusters the values $\delta_4$
and $\beta_4$ seem to be disentangled. For example, large
$\delta_4$ corresponds to small $\beta_4$ in $Na^+_{15}$ and {\it
vice versa} in $Na^+_{27}$. This reflects the fact that
hexadecapole moments are determined by both quadrupole and
hexadecapole terms in (\ref{eq:R}) and, moreover, the role of the
quadrupole term is leading.

The calculations show that in light clusters $Na_{15}^+$,
$Na_{27}^+$ and $Na_{35}^+$ the ground state deformations are
characterized by a deep and distinctive minimum with considerable
energy gain as compared with shape isomers.  The number of isomers
increases and their energy difference shrinks with increasing
system size.  The first isomers in $Na_{53}^+$ and $Na_{119}^+$
are given in the Table I.  They exhibit an opposite sign of
quadrupole deformations as compared to the ground state.  Fig.1
shows, as an example, the energy surface of $Na_{119}^+$, which
has three distinct minima. The second isomer demonstrates
a dominant hexadecapole deformation.

In Figs. 2 and 3, the single-particle spectra of prolate $Na^+_{27}$
and oblate $Na^+_{35}$ are shown and compared with the corresponding
spectra in the spherical limit.  It is well visible how the deformation removes
degeneracies. Note that prolate and oblate clusters exhibit the opposite
assignment of quantum numbers inside subshells.

\section{Separable approach to dynamics}

\subsection{Small amplitude dynamics}

Up to now, we have determined stationary states with
a mean field
$\hat{h}_0(\mathbf{r})$, density $\rho_0(\mathbf{r})$ and
single-particle wave functions
$\psi_{q}(\mathbf{r})$.
We are now proceeding the dynamical Kohn-Sham equations
\begin{equation}\label{eq:TDHF}
  \hat{h}({\bf r},t)\psi_q({\bf r},t)=
  i\frac{\partial}{\partial t} \psi_q({\bf r},t)
  \;.
\end{equation}
The many-body wave functions are composed from the single-particle
wave functions  $\psi_q({\bf r},t)$ as Slater determinants
$\Psi({\bf r}_1,...,{\bf r}_N,t)$.
In the linear regime, when only small oscillations about the
stationary state are considered, one gets in first order perturbation
\begin{eqnarray}
  \Psi (t)&\simeq &\Psi_0+\delta\Psi (t),
  \nonumber\\
  \rho({\bf r},t)&\simeq &\rho_0({\bf r})+\delta\rho({\bf r},t),
  \nonumber\\
  \hat{h}({\bf r},t)&\simeq &
  \hat{h}_0({\bf r})+\delta\hat{h}({\bf r},t)
  \nonumber
\end{eqnarray}
where the response mean-field reads in detail
\begin{equation}\label{eq:del_H}
  \delta\hat{h}({\bf r},t)
  =
  (\frac{\delta U_{xc}}{\delta\rho})_{\rho =\rho_{0}}
  \delta \rho({\bf r},t) + \int d{\bf r}_1\frac{\delta \rho({\bf r}_1,t)}
  {\vert {\bf r} - {\bf r}_1\vert }
  \equiv
  \mbox{Tr}_2\left\{\hat{V}_{\rm res,12}\delta\rho_2\right\}.
\end{equation}
The $\delta\hat{h}({\bf r},t)$ is a purely instantaneous and
local operator by virtue of the local-density approximation
inherent in the Kohn-Sham scheme used here. For harmonic
oscillations it can be written as
\begin{equation} \label{eq:dh}
\delta\hat{h}({\bf r},t)=
\delta\hat{h}^+({\bf r})e^{i\omega t}
+\delta\hat{h}^-({\bf r})e^{-i\omega t}.
\end{equation}
Further, for small-amplitude harmonic
oscillations in the vicinity
of the static ground state $\Psi_0$,
the pertinent ansatz for a perturbation of the
wave function reads \cite{Thoul}
\begin{equation} \label{eq:Thouless}
  \delta\Psi (t)
  =
  \sum_{ph} \left(c^{+}_{ph}e^{i\omega t}
                  +c^-_{ph}e^{-i\omega t}\right)
  \hat{a}_p^{\dagger}\hat{a}^{\mbox{}}_h\Psi_0
  \end{equation}
where  $\hat{a}_p^{\dagger}$ and  $\hat{a}_h$ are creation and
annigilation operators of particle and hole states, respectively.
Insertion of (\ref{eq:dh}) and (\ref{eq:Thouless}) into the
time-dependent Kohn-Sham equation (\ref{eq:TDHF}) with subsequent
linearization and selection of the part $\propto e^{-i\omega t}$
yields finally a standard RPA problem for frequencies $\omega$
and corresponding amplitudes $c^\sigma_{ph}$:
 \begin{equation}
 (\varepsilon_{ph}+\sigma\omega ) c^\sigma_{ph} +
  \langle p|\delta\hat{h}^{\sigma}|h\rangle =0
  \;
\label{eq:fullRPA}
\end{equation}
where $\sigma=\pm$ and
\begin{equation}
  \langle p|\delta\hat{h}^{\sigma}|h\rangle =
  \sum_{p'h'}\left(
   \langle ph'|\hat{V}_{\rm res}|hp'\rangle c^{\sigma}_{p'h'}
   + \langle pp'|\hat{V}_{\rm res}|hh'\rangle c^{-\sigma}_{p'h'}
  \right) \; .
 \label{eq:deltaHfull}
 \end{equation}

\subsection{Separable RPA}

The full RPA, although straightforward, can become very expensive
because the dimension of the matrix of $\hat{V}_{\rm res}$ grows huge in
demanding applications. The separable approach (\ref{eq:separ}) is
designed to reduce the expense of such calculations.
Inserting (\ref{eq:separ}) into the full response (\ref{eq:deltaHfull})
yields
\begin{eqnarray}
  \langle p|\delta\hat{h}^\sigma|h\rangle
  &=&
  \sum_k \bar{\alpha}^\sigma_k \langle p|\hat{Q}_{k}|h\rangle
  \; ,
\label{eq:respsep}\\
  \bar{\alpha}_k^{\sigma}
  &=&
  \sum_{k'}\kappa_{kk'}\int d{\bf r}\,Q_{k'}({\bf r})\delta\rho^\sigma({\bf r})
\nonumber\\
  &=&
  \sum_{k'}\kappa_{kk'}\,
   \sum_{ph}
   \left(
    \langle p|\hat{Q}_{k'}|h\rangle c^\sigma_{ph}
   +
    \langle h|\hat{Q}_{k'}|p\rangle c^{-\sigma}_{ph}\right)
  \; .
\label{eq:alpha}
\end{eqnarray}
Note that the values $\delta\hat{h}^{\sigma}({\bf r})$,
$\bar{\alpha}_k^{\sigma}$ and $\delta\rho^{\sigma}({\bf r})$
become independent of $\sigma$ for purely local separable
operators $\hat{Q}_k({\bf r})$. Such situation takes place when the
motion is driven only by variations of time-even densities. This
is just the case for clusters where the dynamics is determined by
oscillation of time-even density of valence electrons. Then,
\begin{equation}
\label{eq:alpha2}
\bar{\alpha}_k=
  \sum_{k'}\kappa_{kk'}\,
   \sum_{ph}
   \langle p|\hat{Q}_{k'}|h\rangle \left( c^+_{ph}
   +c^{-}_{ph}\right)\; .
\end{equation}
Using (\ref{eq:respsep}),
the RPA equations (\ref{eq:fullRPA}) can be reshuffled into
\begin{equation}
\label{eq:calpha}
  c^\pm_{ph}
  =
  -\frac{1}{2}\frac{\sum_k\bar{\alpha}_k\langle p|\hat{Q}_{k}|h\rangle}
        {\varepsilon_{ph}\pm\omega}
  \; .
\end{equation}
Inserting (\ref{eq:calpha}) into (\ref{eq:alpha2}) yields finally
a system of linear homogenous equation for weights
$\bar{\alpha}_k$:
\begin{equation}
  \sum^K_{k'=1} s_{kk'}(\omega)\bar{\alpha}_{k'}=0
\label{eq:SRPA1}
\end{equation}
where
\begin{equation}
  s_{kk'}(\omega)
  =
   \sum_{ph}\frac{<p\vert \hat{Q}_{k} \vert h>
    <p\vert \hat{Q}_{k'} \vert h>\varepsilon _{ph}}
    {\varepsilon _{ph}^2 - \omega^2} -
            \frac{1}{2\kappa_{kk'}}.
\label{eq:SRPA2}
\end{equation}
In Eq. (\ref{eq:SRPA2}) the sum runs over all particle-hole states
of a given multipolarity.
The condition
\begin{equation} \label{eq:SRPA3}
   \det \mid s(\omega) \mid =0
\end{equation}
gives SRPA dispersion equation for eigenvalues $\omega_j$.

The rank of the SRPA matrix (\ref{eq:SRPA1}) is equal to
the number $K$ of the separable operators, which, as is shown below,
is typically 3-6. So the rank is dramatically less than in the case of
involved RPA methods.  This results in drastic simplification of RPA
calculations.  At the same time, the total number of SRPA roots equals
to the number of particle-hole configurations and every SRPA state is
a superposition of these configurations. The SRPA keeps the important
RPA feature to describe the Landau damping but gets this with much less
computational effort.

It's straightforward to show that
Eqs. (\ref{eq:SRPA1})-(\ref{eq:SRPA3}) can be also obtained with a
generalized schematic RPA Hamiltonian
\begin{equation}\label{eq:Sch_RPA}
  \hat{H}
  =
  \hat{H}_0-\frac{1}{2}\sum_{kk'}\tilde{\kappa}_{kk'}\hat{Q}^{\dag}_{k}
  \hat{Q}_{k'}
\;.
\end{equation}
This can be done by  substituting the Hamiltonian
(\ref{eq:Sch_RPA}) to the standard RPA equations
\begin{equation}
  [\hat{H},\hat{O}^{\dagger}_j]=\omega_j \hat{O}^{\dagger}_j
  \;,\quad
  [\hat{H},\hat{Q}_j]=-\omega_j \hat{O}_j
  \;,\quad
  [\hat{O}_j,\hat{O}^{\dagger}_{j'}]=\delta_{jj'}
\end{equation}
for the creation phonon operator $\hat{O}^{\dagger}_j=\sum_{ph}\left(
c^{-}_{ph}\hat{a}^+_p\hat{a}^{\mbox{}}_h +
c^{+}_{ph}\hat{a}^+_h\hat{a}^{\mbox{}}_p \right)$ and its
hermitian conjugate.
The matrix $[\tilde{\kappa}_{kk'}]$ in (\ref{eq:Sch_RPA})
is inverse to the matrix $[\kappa_{kk'}^{-1}]$ in (\ref{eq:SRPA2}).

Note that relative contributions of operators
$\hat{Q}_k(\mathbf{r})$ to RPA
states are not fitted but regulated self-consistently by the
weights $\bar{\alpha}^{j}_{k}$.
Every RPA state $j$ has its own set of the weights.
The normalization condition
for the particle-hole amplitudes $c^{j\sigma}_{ph}$ and weights
$\bar{\alpha}^{j}_{k}$ can be written through derivatives of
$s_{kk'}(\omega_j)$:
\begin{eqnarray}
\sum_{ph}(c^{j-}_{ph})^2-(c^{j+}_{ph})^2)
&=&
\sum_{kk'}\bar{\alpha}^{j}_{k}\bar{\alpha}^{j}_{k'}
\sum_{ph}\frac{<p\vert \hat{Q}_{k} \vert h>
    <p\vert \hat{Q}_{k'} \vert h>\varepsilon _{ph}\omega_j}
    {(\varepsilon _{ph}^2 - \omega^2_j)^2}
\nonumber\\
&=&\frac{1}{2}
\sum_{kk'}\bar{\alpha}^{j}_{k}\bar{\alpha}^{j}_{k'}
\frac{\partial}{\partial\omega_j}s_{kk'}(\omega_j)=2 .
\label{eq:norm}
\end{eqnarray}

\subsection{Self-consistent prescription for separable operators}

Up to now, the operators $\hat{Q}_k(\mathbf{r})$ and strength matrix
$\kappa_{kk'}$ were not still chosen. Obviously,
the success of SRPA depends on the choice.
The better it is, the smaller
number of the separable operators we need to reproduce
$\hat{V}_{\rm res}$. We are considering systems with pronounced
collective modes (the various plasmons in case of clusters) and so
just collective deformations should motivate the form of
$\hat{Q}_k(\mathbf{r})$. Here we will follow the VPM
prescription \cite{Lip} which was very effective
in the SRPA calculations
for spherical clusters \cite{Ne_EPJ}. In the VPM, the operators
$\hat{Q}_k(\mathbf{r})$ are self-consistently generated by
collective variations of the density which in turn is determined
by the scaling transformation \cite{Lip,sumrul}. Being thus constructed,
only few operators are enough to cover successfully the dynamics.
Specifically, the collective deformation is achieved by the scaling
transformation
\begin{equation} \label{eq:scale}
  \Psi (t)  =
  \prod_{k=1}^K
  e^{\alpha_{k}(t) \sum^{N_e}_{i=1}
  [\hat{h}_0({\bf r}_i),f_{k}({\bf r}_i)]}
  \Psi_0
  \;.
\end{equation}
Here, the local operators $f_{k}({\bf r})$ determine the intended time-even
shape of the deformation. For example, $f=z$ leads to a translation along
$z$-direction, $f=r^2Y_{20}$ results in an ellipsoidal deformation. The
collective amplitudes $\alpha_{k}(t)=\tilde{\alpha}_{k}cos\,\omega t$
determine the relative weights of different deformation modes.
For small deformations $\tilde\alpha_k$, the density variation
becomes
\begin{eqnarray}\label{eq:del_rho}
 \delta \rho({\bf r},t)
 &=&
 (\Psi (t)|\hat{\rho}|\Psi (t))-(\Psi_0|\hat{\rho}|\Psi_0)\nonumber\\
 &\simeq&
 \sum^K_{k=1} \alpha_{k}(t)    
 ({\bf\bigtriangledown} \rho_{0}({\bf r}) \cdot
 {\bf\bigtriangledown} f_{k}({\bf r}) +\rho_{0}({\bf r})
 \triangle f_{k}({\bf r}))
\end{eqnarray}
which, after submission into (\ref{eq:del_H}), results in the
response Hamiltonian
\begin{equation}
  \delta \hat{h}({\bf r})=                                            
  \sum^K_{k=1} \alpha_{k} \hat{Q}_{k}({\bf r})
\label{eq:respcoll}
\end{equation}
composed from the operators
\begin{equation}
\label{eq:Q}
  \hat{Q}_{k}({\bf r})
  =
  \int d{\bf r}_1 V_{\rm res}({\bf r}-{\bf r}_1)
  \left(\bigtriangledown \rho_0({\bf r}_1)\cdot
              \bigtriangledown f_{k}({\bf r}_1)
       + \rho_0({\bf r}_1) \triangle f_{k}({\bf r}_1)\right)
\end{equation}
with
\begin{equation}
  V_{\rm res}({\bf r}-{\bf r}_1)
  =
  (\frac{\partial U_{xc}}{\partial\rho})_{\rho =\rho_{0}}
  \delta({\bf r}-{\bf r}_1)
  +
  \frac{e^2}{\vert {\bf r} - {\bf r_1}\vert }
  \;.
\nonumber
\end{equation}
Operators (\ref{eq:Q}) take, by construction, into account all
deformation distortions of the ground state (this point is discussed
in the Appendix B where the explicit expression
for $\hat{Q}_k(\mathbf{r})$ in deformed clusters is given.)
It is now self-suggesting that these operators  are identical with
the basis operators to be used in the separable ansatz
(\ref{eq:separ}). This amounts to identify the collective
response (\ref{eq:respcoll}) of  operators (\ref{eq:Q})
with that from the separable ansatz (\ref{eq:respsep}),
which finally yields the strengths coefficients as
\begin{equation}
\label{eq:kappa}
\kappa^{-1}_{kk'} =
           - \int d{\bf r} Q_{k}({\bf r})
             (\bigtriangledown \rho_0 ({\bf r})\cdot
             \bigtriangledown f_{k'}({\bf r})
         + \rho_0({\bf r}) \triangle f_{k'}({\bf r}))\; .
\end{equation}
Eqs. (\ref{eq:calpha})-(\ref{eq:Sch_RPA}), (\ref{eq:norm}), and
(\ref{eq:Q})-(\ref{eq:kappa})
constitute the complete set of {\it self-consistent} SRPA
equations. One has to point out that, like involved RPA,
SRPA maintains full details of $1ph$ structure in the spectra.
The only point is that the residual interaction is parametrized
in terms of collective picture.

It is easy to see from (\ref{eq:scale}) that just input local
operators $f_{k}({\bf r})$ determine the character of the
perturbation. The choice of the operators is the most delicate point
of the approach and devotes a special attention (see discussion in
Appendix C). We use the set of hermitian operators:
\begin{equation} \label{eq:lop}
f_{\lambda_k p_k\mu}({\bf r})=r^{\lambda_k + p_k}
(Y_{\lambda_k\mu}(\Omega)+                                  
Y_{\lambda_k\mu}^{\dag}(\Omega))
\end{equation}
with
\begin{eqnarray} \label{eq:lopspec}
\lambda_k p_k =10,12,14,30,32  \mbox{ for }
(\mu =0,1), \nonumber \\
\lambda_k p_k =20,22,24,40,42 \mbox{ for }
 (\mu =0,1,2),\\
\lambda_k p_k =30, 32, 34, 50, 52 \mbox{ for }
(\mu =0,1,2,3) \nonumber
\end{eqnarray}
for dipole, quadrupole and octupole excitations, respectively.
It is seen that in all the sets the first operator
has the form of the applied electric external field in the
long-wave approximation, the others two represent a variety of the
radial dependencies to cover different slices of the system and the last
ones take into account the coupling of different $\lambda$-modes with equal
$\mu$ and parity, which takes place in deformed systems.
Relative contributions of different input operators $f_{\lambda_k
p_k\mu}({\bf r})$ to the residual interaction are self-consistently regulated
by the strict computation of the strengths according
to Eq. (\ref{eq:kappa}).

\subsection{Response to electric fields}

The photoabsorption cross section for excitation
of the state $j=\Lambda^{\pi}$
in axially deformed cluster has the form
\begin{equation}\label{eq:sig}
\sigma (E\lambda\mu ; gr \mapsto j)=
8\pi^3\frac{\lambda +1}{\lambda [(2\lambda +1)!!]^2}
(\frac{\omega_j}{\hbar c})^{2\lambda -1}
\delta_{(-1)^{\lambda},\pi}\delta_{\mu ,\Lambda}
M_{\lambda\mu j}^2
\end{equation}
where
\begin{eqnarray} \label{eq:M}
  M_{\lambda\mu j}
  &=&
  <j|r^{\lambda}Y_{\lambda\mu}|0>
\\
  &=&
  \frac{1}{\sqrt{2}}\sum_{ph}<p|r^{\lambda}Y_{\lambda\mu}|h>
   (c^{j+}_{ph}+c^{j-}_{ph})
  =
  -\frac{1}{\sqrt{2}}\sum_k \bar{\alpha}_{k}^{j} A_k(\omega_j)
\nonumber
\end{eqnarray}
is the reduced matrix element of $E\lambda\mu$ transition
from the ground state to the RPA state $j$ and
\begin{equation}
A_{k}(\omega_j)=\sum_{ph} \frac{\varepsilon_{ph}
<p|r^{\lambda}Y_{\lambda\mu}|h><p|\hat{Q}_{k}|h>}
{\varepsilon_{ph}^2-\omega_j^2}\;.
\end{equation}

Plasmons in deformed clusters involve a lot of RPA states
which in any case cannot be well resolved. So, it worth to
consider not every RPA state but an averaged response.
Strength function formalism allows to get the averaging
response without dealing with RPA problem for every state,
which gives a huge gain in the computational effort.
The factorized residual interaction used in SRPA makes
derivation of the strength function especially simple
\cite{TMF}. Following this line, the strength function
is defined as
\begin{equation} \label{eq:sig_av}
\sigma (E\lambda\mu ,\omega )=\sum_j
\sigma (E\lambda\mu ; gr \mapsto j) \eta (\omega -\omega_j)
\end{equation}
where
\begin{equation}
\eta(\omega - \omega_j) = \frac{1}{2\pi}
  \frac{\Delta}{(\omega - \omega_j)^2 + (\Delta/2)^2}              
\label{eq:lorfold}
\end{equation}
is Lorentz weight with an averaging parameter $\Delta$.
Using Cauchy residue theorem, one may rewrite (\ref{eq:sig_av})
in the form independent of $j$ (when the response is directly computed
for given $\omega$). The derivation starts with the squared
matrix element of $E\lambda\mu$
transition which, provided that SRPA matrix $s_{kk'}$
(Eq. (\ref{eq:SRPA2})) is symmetric, can be written as
\begin{equation}
M_{\lambda\mu j}^2=\frac{G(\omega_j)}{\frac{\partial}
{\partial\omega_j} det|s(\omega_j)|}
\end{equation}
where
\begin{equation}
G(\omega_j)=\sum_{kk'}S_{kk'}(\omega_j)A_k(\omega_j)A_{k'}(\omega_j),
\end{equation}
and $S_{kk'}(\omega_j)$ is the algebraic supplement of the
matrix element $s_{kk'}(\omega_j)$. Then, provided that
the function $1/\det |s|$ has only one-multiple poles
$\omega=\pm\omega_j$, the strength function
(\ref{eq:sig_av}) is expressed through the corresponding
residues on the complex plain z:
\begin{equation}
\sigma (E\lambda\mu ,\omega ) =8\pi^3\frac{\lambda +1}{\lambda
[(2\lambda +1)!!]^2} \frac{1}{(\hbar c)^{2\lambda -1}}\sum_j Res
[F(\omega ,z)]_{z=\pm\omega_j}
\end{equation}
where
\begin{equation}
\label{eq:F}
F(\omega ,z)=\frac{z^{2\lambda -1}\eta (\omega -z)G(z)}{\det|s(z)|}
\; .
\end{equation}

Following Cauchy theorem, sum of all the residues (covering all
possible poles of the function (\ref{eq:F})) is zero
and so we can express the residues with $z=\omega_j$ through
the rest of the others:
\begin{eqnarray}
Res[F]_{z=\omega_j}=&-&(
Res[F]_{z=-\omega_j}+Res[F]_{z\to\infty}
\nonumber\\
&+&Res[F]_{z=\omega \pm i(\Delta /2)}+Res[F]_{z=\pm\varepsilon_{ph}}
) \;.
\end{eqnarray}
It's easy to prove that $\lim_{|z|\to \infty}F(\omega ,z)=0$
for $\lambda =1,2$. Also, $Res[F]_{z=-\omega_j}$ and $Res[F]_{z=-\varepsilon_{ph}}$
can be neglected for large positive $z$-values in the
plasmon energy region and for relevant values of
the averaging parameter $\Delta$. Remaining residues over the poles
$z=\omega \pm i(\Delta /2)$ and $z=\varepsilon_{ph}$
give the final outcome
\begin{equation}
\label{eq:sf}
\sigma (E\lambda\mu ,\omega )
=8\pi^3\frac{\lambda +1}{\lambda [(2\lambda +1)!!]^2}
\frac{1}{(\hbar c)^{2\lambda -1}}
\end{equation}
$$
\times
\left(
\frac{-1}{\pi} \Im
\left[
\frac{z^{2\lambda -1}G(z)}{\det|s(z)|}
\right]_{z=\omega +i(\Delta /2)}
+\sum_{ph}\varepsilon_{ph}^{2\lambda -1}
<p|r^{\lambda}Y_{\lambda\mu}|h>^2\eta(\omega -\varepsilon_{ph})
\right) .
$$
Unlike expression for the standard strength
function \cite{Ring,Brack}, Eq. (\ref{eq:sf}) consists from two terms.
The first one is contribution from the residual interaction
and the second one is unperturbed strength function.
This form is convenient for the analysis. The standard strength
function can be reduced to the form (\ref{eq:sf}) if
the width $\Delta$ is introduced to
$\delta (\omega\pm\omega_j)$-distribution. This gives
the Lorentz shape.
Eq. (\ref{eq:sf}) can be used for calculation of E1 and E2
photoabsorption.

Model-independent energy-weighted sum rules (EWSR)
\begin{eqnarray}\label{eq:EWSR}
S(E\lambda )&=& \sum_{\mu}\sum_{ph}\varepsilon_{ph}
<p|r^{\lambda}Y_{\lambda\mu}|h>^2
=\sum_{\mu}\sum_{j} \omega_j M_{\lambda\mu j}^2
\nonumber\\
&=&\frac{\hbar^2 e^2}{8\pi m_e} \lambda (2\lambda + 1)^2 N_e
<r^{2\lambda -2}>
\end{eqnarray}
is used to control the completeness of the 1ph configuration
space.

\section{Results and discussion}

\subsection{Dipole plasmon}

Results of SRPA calculations for the dipole plasmon are exhibited in
Figs. 4-6. Two different presentations are used. First, to demonstrate
the complex structure of the plasmon and Landau damping, the
photoabsorption strength (\ref{eq:sig}) is given for every RPA state
by a vertical bar (in $eV\AA$). Second, for the convenience of
the comparison with experimental data, the strength function
(\ref{eq:sig_av}) averaged with the Lorentz weight is used. The value of
the averaging parameter is $\Delta =0.25$ eV to simulate the typical
thermal broadening of the plasmon.

In Fig. 4, the photoabsorption is presented for three light deformed
clusters, prolate $Na^+_{15}$ and $Na^+_{27}$ and oblate $Na^+_{35}$.
The SRPA calculations are compared with recent experimental data
\cite{exp} obtained at rather low temperature $T=105 K$.  The
agreement with experiment is very satisfying.  It should be mentioned
that the average energy of the plasmon was slightly corrected by
decrease of the diffuseness parameter $\alpha$ in (\ref{eq:densi})
from 1.0 to 0.8 ${\rm a}_0$. The value 1.0 ${\rm a}_0$ chosen to
simulate ionic effects at high temperatures ($T=400-800
K$) \cite{LRPA2} is not suitable for $T=105 K$.  The decrease of the
surface diffuseness $\alpha$ leads to a blueshift of the plasmon,
which is indeed seen when lowering the temperature.  Beside the good
description of the average energy, the deformation splitting of the
plasmon into two peaks is well reproduced.  The peaks correspond to
$\mu =0$ and 1 dipole modes.  The latter involves contributions from
both $\mu =\pm 1$ projections and so is about twice as large. This
mode forms the upper part of the plasmon spectrum in prolate clusters
and the lower part in oblate ones (see also Fig. 9 e)).  The splitting
is well developed in the prolate clusters and less strong in the
oblate $Na^+_{35}$. Though first evidence of the Landau damping is
already seen, the plasmon gross-structure and width are still mainly
determined by the deformation splitting.

Following such a splitting analysis, one is tempted to associate with
$Na^+_{53}$ (see Fig. 5) an oblate deformation.  Its plasmon shape is
rather similar to the one for oblate $Na^+_{35}$ and the broad peak
with a small right shoulder at 3.1 eV can, in principle, be
approximated by two Lorentz curves ``justifying'' an oblate shape, as
done e.g. in \cite{exp}.  However, our calculations show that
this interpretation is misleading. Unlike lighter clusters,
$Na^+_{53}$ shows considerable Landau damping (see bars in Fig. 5, top
panel). This broadens the plasmon and modifies its gross-structure. In
particular, the structure at 3.1 eV is caused not by $\mu= 0$ mode but
by the group of $\mu =1$-states.  Our calculations predict for this
cluster the prolate shape, which results in very satisfactory
agreement with the experimental data. Good description is also
obtained for other clusters in this size region (see the
comprehensive SRPA analysis of the data \cite{exp} in
\cite{NKR}). Our conclusion on prolate shape of clusters in this size
region agrees with calculations of other groups \cite{Thomas}.
Altogether this means, that generally accepted experimental practice
to treat plasmon shape within simple deformation models and thus
to conclude on cluster deformation is not appropriate
for medium-size clusters. This way does not take into account
important role of Landau damping in forming plasmon gross-structure
and width and so can be misleading.

For medium and large clusters, the disentangling of the dipole
spectrum becomes even more complicated because of possible
incoherent contributions from isomers.  Table I shows that
$Na^+_{53}$ has oblate isomeric state with a tiny energy
deficit as compared with the ground state. This isomer can also
contribute to the observed spectrum. The bottom panel of Fig. 5
demonstrates that the isomeric state yields similar pattern, in
spite of reverse ordering of $\mu =0$ and 1 modes. In
$Na^+_{53}$ both the modes are broad and have two-peak
structure. In some important details the isomer plasmon deviates
from the experimental data.

The picture is complicated with further growing cluster
size. As was shown in Fig.1, the energy surface in $Na^+_{119}$ has
already three distinctive minima: oblate ground state, first prolate
isomer and second hexadecapole isomer. The corresponding plasmons are
exhibited in Fig. 6. Contributions of $\mu =0$ and 1 modes are also
given to show the role of the deformation. All the plasmons
demonstrate strong Landau damping which determines, in a large extent,
their width. Just because of Landau damping, the plasmons,
in spite of different cluster deformations, show very similar
broad shapes. Obviously, all three deformation
configurations of $Na^+_{119}$ can contribute to the
measured spectrum. Indeed, clusters exhibit thermal shape fluctuations
which are incoherently added in experiment. The shapes of the ground
state and first isomers have to provide the dominant contribution.
The observed plasmon reflects a statistical mix of many shapes with a few
dominants.

To summarize, the calculations demonstrate increasing importance
of Landau damping and isomer mixing for the dipole spectrum.  Already
at medium cluster sizes these factors cannot be neglected.  As a
result, the often used prescription to conclude on cluster shape on
the grounds of approximation of plasmon shape by 1-3 Lorentz curves
is insufficient for these system sizes.

The strong fragmentation of the spectra through Landau damping may
raise doubts on the collectivity of the plasmon oscillation.  The
strength is distributed over many sub-states and calculations show
that even the strongest RPA states do not exceed more than 2-4 times
the strength of particle-hole configurations, in heavy clusters even
less.  Nevertheless, the entity of RPA states near the plasmon
frequency can still be called collective. One may view this as a two
step process. The dipole excitation first generates a collective
surface plasmon state which is then quickly distributed over the nearby
particle-hole states through Landau damping. The lifetime of the
collective dipole excitation can be estimated to be about 10 fs
in large Na clusters \cite{ownrev}. The  collectivity is also
testified by reach particle-hole structure of RPA states.

\subsection{Quadrupole and octupole plasmons}

In Figs. 7-8 the photoabsorption cross sections
for quadrupole and octupole plasmons in prolate $Na^+_{27}$ and
oblate $Na^+_{35}$ are presented. Contributions from all the
projections $\mu$ are given to demonstrate the role of the
deformation broadening.

The figures show strong Landau damping. Its contribution to the width
and gross structure of plasmons is of the same scale or even larger
than from deformation splitting.  The prolate $Na^+_{27}$ has a strong
deformation and we see a distinctive blueshift of $\mu$-branches of
the plasmon with increasing $\mu$. This makes the deformation
splitting comparable with Landau damping.  The deformation is weaker
in the oblate $Na^+_{35}$, which makes the deformation splitting
less pronounced than Landau damping.

Both, quadrupole and octupole plasmons, are well concentrated
(especially in the less deformed $Na^+_{35}$), which helps their
experimental observation. The plasmons have a high-energy tail.
Analysis of structure of main RPA states from the
plasmon region shows that they have about the same character as in the
dipole case.  There is initially a collective multipole plasmon
which is then quickly Landau damped and fragmented over the many
nearby particle-hole states.

As is seen from the figures, photoabsorption cross sections for
quadrupole and octupole plasmons are $10^5$ and $10^{10}$ times weaker
than in the dipole case. Both plasmons share the energy region with
the all dominating dipole plasmon and so should be completely masked by the
latter. Strong Landau damping of the plasmons additionally decreases
their chance to be observed in photoabsorption. However, following the
estimates \cite{Sol_ee}, inelastic electron scattering could resolve
these states.  At certain angles of
outcoming electron, contributions of $\lambda =2$ and 3 plasmons to
the cross section can successfully compete with the dipole
contribution. The reaction is very demanding as it requires an
intense, monochromatic and well collimated low-energy electron
beam. However, progress of experimental techniques gives us some hope.
SRPA results can be used in such experiments as a first guide for
plasmon energies and strength distributions.

\section{Conclusions}

The separable RPA (SRPA), a self-consistent approach to RPA, is
presented in connection with investigation of multipole plasmons in
deformed clusters.  Both ground state properties and RPA response are
calculated using the same Kohn-Sham functional for axial shapes. Due
to self-consistent factorization of the residual interaction,
SRPA dramatically simplifies RPA calculations. At the same time,
the method ensures high accuracy of the calculations. As compared with other
RPA methods, SRPA provides a comprehensive treatment of Landau
damping with a {\it minimal computational effort}. This is crucial
for deformed systems
where one deals with an impressive particle-hole configuration space.

SRPA has been applied to dipole, quadrupole and octupole plasmons in a
variety of axial clusters. The main attention was paid to the dipole
plasmon.  Light, medium and heavy clusters of both prolate and oblate
shapes were covered. Both quadrupole and hexadecapole deformations
were taken into account. Contributions to the plasmon from both ground
and isomeric states were analyzed.  Good agreement with available
experimental data was achieved.

Unlike light axial clusters, where dipole plasmon has a typical
two-peak structure reflecting the deformation splitting, the dipole
spectra in medium and heavy deformed clusters show usually a broad
bump with relatively small variations of the shape.  The deformation
splitting is masked by strong Landau damping which inhibits a direct
determination of ground state deformations from the spectra.
Moreover, such clusters have shape isomers which further overlay the
spectra. As a result, a correct treatment of plasmon properties is
possible only on the grounds of fully microscopic calculations taking
into account all these effects. SRPA is a promising method for such
investigations.

Quadrupole and octupole spectra were shown to stay well concentrated
in spite of the deformation splitting and Landau damping.  Both
spectra have a considerable high-energy tail.  The quadrupole and
octupole spectra cannot be observed directly in photoabsorption
experiments but there is some hope to disentangle them in inelastic
electron scattering.

\newpage
{\bf \large Appendix A}
\vspace{0.2cm}

The problem (\ref{eq:H0}) for the deformed Kohn-Sham potential
\begin{equation} \label{eq:deKS}
(-\frac{\hbar^2}{2m}\bigtriangledown^2 +U_{C}({\bf
r})+U_{xc}({\bf r}))\psi_{q}=\varepsilon_{q} \psi_{q}
\end{equation}
can be rewritten as a system of equations for ampltitudes
$a^{q}_{s}$ of the expansion (\ref{eq:psi}):
\begin{equation} \label{eq:eq1}
(E_{s}-\varepsilon_{q})a^{q}_{s}+ \sum_{q '}a^{q
'}_{s}<\phi_{s}|{\tilde U}_{C}({\bf r})+{\tilde U}_{xc}({\bf
r})|\phi_{s'}>=0
\end{equation}
where
\begin{equation} \label{eq:eq2}
{\tilde U}_{C}({\bf r})=U_{C}({\bf r})-U_{C}^0({\bf r}),
\end{equation}
\begin{equation} \label{eq:eq3}
{\tilde U}_{xc}({\bf r})=U_{xc}({\bf r})-U_{xc}^0({\bf r}),
\end{equation}
and values $U_{C}^0({\bf r})$ and $U_{xc}^0({\bf r})$ enter the
problem for spherical Kohn-Sham potential
\begin{equation} \label{eq:sKS}
(-\frac{\hbar^2}{2m}\bigtriangledown^2 +U_{C}^0({\bf
r})+U_{xc}^0({\bf r}))\phi_{s}=E_{s} \phi_{s}.
\end{equation}
For the quantities (\ref{eq:densi}), (\ref{eq:dedens}),
(\ref{eq:eq2}), and (\ref{eq:eq3}) the multipole expansion
is used.
In particular, the electron ground state density (\ref{eq:dedens})
is written as (see notations in Sec. 2)
\begin{equation} \label{eq:dedens1}
\rho_0 ({\bf r})=\sum_{{\tilde \lambda}} Y_{{\tilde \lambda}0}
(\Omega) {\tilde\rho}_{{\tilde \lambda}}(r),
\end{equation}
with
\begin{eqnarray}
\label{eq:dedens2}
{\tilde\rho}_{{\tilde \lambda}}(r)&=&\frac{1}{\sqrt{\pi (2{\tilde
\lambda}+1)}} \sum_{q={\cal N}n_z\Lambda}^{occ} (-1)^{\Lambda}
(2-\delta_{\Lambda ,0}) w_q
\nonumber\\
&\times &\sum_{n_2, l_2, n_1,l_1}
\sqrt{(2l_2+1)(2l_1+1)} C^{{\tilde\lambda},0}_{l_2,0, l_1,0}
C^{{\tilde\lambda},0}_{l_2,-\Lambda, l_1,\Lambda}
\\
&\times &a^{q}_{n_2l_2}a^{q}_{n_1l_1}
 R_{n_2l_2}(r)  R_{n_1 l_1}(r). \nonumber
\end{eqnarray}
In all the multipole expansions, including Eq. (\ref{eq:dedens1})
we take ${\tilde \lambda}=0,2,4,6,8$, which provides
sufficient numerical accuracy.

Kohn-Sham equations (\ref{eq:eq1}) together with the density
equation (\ref{eq:dedens1}) are solved by an iteration method.
Eigenvalues $\varepsilon_{s}$ and eigenfunctions $\phi_{s}$ for
the spherical Kohn-Sham potential are used as the input.

The equilibrium deformations $\delta_2$ and $\delta_4$ are
determined by minimization of the total energy (\ref{eq:func}).
It is convenient to use for the kinetic energy  the expression
\begin{eqnarray}
E_T&=&\frac{\hbar^2}{m_e}\sum_{q={\cal N}n_z\Lambda}^{occ}(2-\delta_{\Lambda,0})\sum_{n_1,n_2}\sum_l
a^{q}_{n_2l}a^{q}_{n_1l}
\\
&\times & (\int {dr} r^2\frac{dR_{n_1l}(r)}{dr} \frac{dR_{n_2l}(r)}
{dr} + l(l+1)\int dr R_{n_1l}(r) R_{n_2 l}(r)). \nonumber
\end{eqnarray}
where $(-1)^l=\pi$, $\Lambda\ge 0$.
Further \cite{GL},
\begin{equation}
 E_{xc}=\int d{\bf r}(\epsilon_{xc}(\rho_0 ({\bf r}))
 \rho_0 ({\bf r})
\end{equation}
with
\begin{eqnarray}
\epsilon_{xc}&=&
\epsilon_{x}+ \epsilon_{c}
\;,
\\
\epsilon_{x}&=&\frac{0.916}{r_{s}}Ry\; a_0, \quad
r_{s}(\rho_0)=(\frac{3}{4\pi \rho_0})^{1/3},
\nonumber
\\
\epsilon_{c}&=&-0.066 Ry
\{(1+x^3)log(1+x^{-1})+\frac{1}{2}x-x^2-1/3\},
\nonumber
\\
x&=&\frac{r_{s}}{11.4\; a_0}\;.
\nonumber
\end{eqnarray}
The Coulomb term can be written as
\begin{equation}
E_C=\frac{e^2}{2}\int d{\bf r}(\rho_0({\bf r}) - \rho_i({\bf r}))U_C({\bf
r})\; .
\end{equation}

\newpage
{\bf \large Appendix B}
\vspace{0.2cm}

Here we present explicit expression for the
operator of residual interaction (\ref{eq:Q})
in deformed clusters.
Making use a multipole expansions for densities (\ref{eq:densi}) and
(\ref{eq:dedens}) and potentials (\ref{eq:eq2}) and (\ref{eq:eq3}),
one gets
\begin{equation}
\label{eq:Qdef}
Q_{k}({\bf r})=
\sum_{\tilde \lambda}
\sum_{L} (Y_{L\mu}(\Omega)+Y_{L\mu}^{\dagger}(\Omega))
\nonumber
\end{equation}
$$
 \times[C^{L\mu}_{{\tilde \lambda}0\lambda_k\mu}
 Q^{(C)}_{kL{\tilde \lambda}}(r)
 +\sum_{\tilde{\lambda}'}\sum_{L'}
 (Y_{L'\mu}(\Omega)+Y_{L'\mu}^{\dagger}(\Omega))
 C^{L'\mu}_{{\tilde \lambda'}0L\mu}
 Q^{(xc)}_{kLL'{\tilde \lambda}{\tilde \lambda}'}(r)]
$$
where
\begin{eqnarray}
Q^{(C)}_{kL{\tilde \lambda}}(r) &=& -\frac{2}{\sqrt{4\pi(2L+1)}}
\frac{4\pi e^2}{2L+1}
\nonumber
\\
&\times &\lbrack r^{-(L+1)}
\int \limits_{0}^{r} dr_1 {\cal R}_{L\tilde{\lambda}}^{(k)}(r_1) r_1^{L+2}
+ r^L \int \limits_{r}^{\infty} dr_1
{\cal R}_{L\tilde{\lambda}}^{(k)}(r_1) r_1^{-(L-1)}
  \rbrack\; ,
\nonumber
\\
Q^{(xc)}_{kLL'{\tilde \lambda}{\tilde \lambda}'}(r)&=& -\frac{1}{2\pi}
\sqrt{\frac{2{\tilde \lambda}'+1}{2L'+1}}
C^{L'0}_{L0{\tilde \lambda}'0}
{\cal R}_{L\tilde{\lambda}}^{(k)}(r)
U_{xc}^{\tilde{\lambda}'}(r)\;,
\nonumber
\\
{\cal R}_{L\tilde{\lambda}}^{(k)}(r)&=& \frac{1}{2}r^{p_k-1}
 \lbrace \frac{d\rho_{\tilde{\lambda}}(r)}{dr}
 N_{L\tilde{\lambda}}^{(k)}
+ \frac{\rho_{\tilde{\lambda}}(r)}{r}
M_{L\tilde{\lambda}}^{(k)}
\rbrace\;.
\nonumber
\end{eqnarray}
Here, $\rho_{\tilde{\lambda}}(r)$ and $\tilde{U}_{xc}^{\tilde{\lambda}}(r)$
are expansion coefficients  in (\ref{eq:dedens2}) and
$U_{xc}({\bf r})=\sum_{\tilde{\lambda}} U_{xc}^{\tilde{\lambda}}(r)
Y_{{\tilde \lambda}0}(\Omega)$, respectively, and
\begin{eqnarray}
N_{L\tilde{\lambda}}^{(k)}&=&(p_k + \lambda_k + 1) \sqrt{\lambda_k (2\lambda_k -1)}
(B^{(1)}_{L\tilde{\lambda}k}-A^{(1)}_{L\tilde{\lambda}k})
\nonumber
\\
&-&(\lambda_k -p_k) \sqrt{(\lambda_k +1)(2\lambda_k +3)}
(B^{(2)}_{L\tilde{\lambda}k}-A^{(2)}_{L\tilde{\lambda}k})\;,
\nonumber
\\
M_{L\tilde{\lambda}}^{(k)}&=&(p_k + \lambda_k + 1) \sqrt{\lambda_k(2\lambda_k -1)}
((\tilde{\lambda}+1) B^{(1)}_{L\tilde{\lambda}k} + l A^{(1)}_{L\tilde{\lambda}k})
\nonumber
\\
&+&(\lambda_k -p_k)\sqrt{(\lambda_k +1)(2\lambda_k +3)}
((\tilde{\lambda}+1) B^{(2)}_{L\tilde{\lambda}k} + l A^{(2)}_{L\tilde{\lambda}k})
\nonumber
\\
&+&(\lambda_k -p_k)(p_k +\lambda_k +1) \sqrt{(2\lambda_k +1)(2\tilde{\lambda} +1)}
C^{L0}_{\tilde{\lambda}0 \lambda_k0}
\nonumber
\end{eqnarray}
where
$$
A^{(1)}_{L\tilde{\lambda}k}=
\sqrt{(\tilde{\lambda}+1)(2\tilde{\lambda}+3)}
\pmatrix{
\tilde{\lambda}+1     & \lambda_k-1 & L \cr
\lambda_k & \tilde{\lambda}         & 1 \cr}
 C^{L0}_{\tilde{\lambda}+1\; 0\; \lambda_k-1\; 0}\;,
$$
$$
A^{(2)}_{L\tilde{\lambda}k} = \sqrt{(\tilde{\lambda}+1)(2\tilde{\lambda}+3)}
\pmatrix{
\tilde{\lambda}+1     & \lambda_k +1 & L \cr
\lambda_k & \tilde{\lambda}         & 1 \cr}
 C^{L0}_{\tilde{\lambda}+1\; 0 \;\lambda_k+1\; 0}\;,
$$
$$
B^{(1)}_{L\tilde{\lambda}k} = \sqrt{\tilde{\lambda}(2\tilde{\lambda}-1)}
\pmatrix{
\tilde{\lambda}-1     & \lambda_k-1 & L \cr
\lambda_k & \tilde{\lambda}         & 1 \cr}
 C^{L0}_{\tilde{\lambda}-1\; 0\; \lambda_k-1\; 0}\;,
$$
$$
B^{(2)}_{L\tilde{\lambda}k} = \sqrt{\tilde{\lambda}(2\tilde{\lambda}-1)}
\pmatrix{
\tilde{\lambda}-1     & \lambda_k+1 & L \cr
\lambda_k & \tilde{\lambda}         & 1 \cr}
 C^{L0}_{\tilde{\lambda}-1\; 0\; \lambda_k+1\; 0}\;.
$$
The values $C^{L\Lambda}_{{\tilde \lambda}0\lambda_k\Lambda}$ and those
in big parentheses are Clebsch-Gordan coefficients and $6j$-symbols, respectively.

Expression (\ref{eq:Qdef}) shows that coupling of
oscillations of given multipolarity ($\lambda_k\mu$)
with the spherical $({\tilde \lambda} =0)$ and
deformed $({\tilde \lambda}=2,4,..)$ parts of the
density leads to a family of terms with
$\mid {\tilde \lambda}-\lambda_k \mid \leq L \leq {\tilde \lambda}+\lambda_k$.
Thus the residual interaction driven by the operator (\ref{eq:Qdef})
takes into account not only the deformation distortions of the ground state density
but also the coupling of $\mu^{\pi}$ modes with different
$\lambda$. This coupling is pertinent only to deformed systems
(see, e.g. discussion in \cite{Ne_PRL}).

\newpage
{\bf \large Appendix C}
\vspace{0.2cm}

Here we will comment some essential points concerning the scaling
(\ref{eq:scale}), choice of  input operators
$f_{\lambda_k p_k\mu}({\bf r})$, and convergence of the SRPA results.

 Time-odd hermitian generators
${\hat P}_{\lambda_k p_k\mu}({\bf r})=
-i[\hat{h}_0({\bf r}),f_{\lambda_k p_k\mu}({\bf r})]$ entering the scaling
result in time-even paths $\alpha_{k}(t)$. The generators can be
considered as momenta conjugate to time-even hermitian coordinate-like
operators $f_{\lambda_kp_k\mu}({\bf r})$. Since in our case
all the dynamics is provided by the time-even electron density, there is no
any need in time-even generators. This explains why the scaling
includes not full but single-particle Hamiltonian. Indeed, if
separable residual interaction in a schematic full Hamiltonian
involves only time-even hermitian $\hat{Q}_k({\bf r})$-operators, then
\begin{eqnarray}
[\hat{H},f]_{ph} \simeq [\hat{h}_0,f]&-&\sum_{kk'} \kappa_{kk'}
(\hat{Q}_{k}<\Psi_0 |[\hat{Q}_{k'},f]|\Psi_0>
\nonumber\\
&+&
<\Psi_0 |[\hat{Q}_{k},f]|\Psi_0>\hat{Q}_{k'})
=[\hat{h}_0,f]\; ,
\end{eqnarray}
since averaged
commutators between hermitian operators of the same
time-parity is zero.

It is easy to see from (\ref{eq:scale}) that collective motion
is determined by input
operators $f_{\lambda_k p_k\mu}({\bf r})$.
The choice of the operators is motivated by physical reasons.
The number of the operators should be as small as possible to
minimize the computational effort and, at the same time,
should be sufficient  to ensure the convergence of the results.
Specifically, if we study response of the system
to an external field, then one of the operators must take the form
of such field. One input operator is usually not enough to reproduce
the true residual interaction, even if the separable interaction
is constructed self-consistently. To overcome this trouble, we exploit
the idea of the local RPA \cite{LRPA1,LRPA2} to use not one but several
input operators. In the local RPA the cluster is treated
as a system of several coupled oscillators, which allows to describe
the gross structure of the resonance. Instead, SRPA
can be viewed as a system of coupled schematic RPAs. The operators
$f_{\lambda_kp_k\mu}({\bf r})$ are chosen so that the corresponding
operators  of the residual interaction, $\hat{Q}_k$, have maxima
at different
slices of the system, including both surface and interior.
Such input operators can
be intuitively associated with different external probes
causing  the response of different parts of the system. Besides,
several operators allow the system to relax additionally, which
finally leads to high accuracy of the method \cite{Ne_EPJ}.
The number of input operators is dictated by
convergence of the results and usually 3-6 operators are enough.

Fig. 9 demonstrates the relative role of different input
operators in forming the dipole plasmon. Prolate
$Na_{27}^+$ and oblate $Na_{35}^+$ are considered as typical examples.
It is seen that the residual interaction posed by the
leading input operator $rY_{1\mu}$  shifts the unperturbed dipole
strength from 1-1.5 eV to the energy $\sim 2.5$ eV. This operator
is enough to get the correct excitation energy. However,
there still exists an artificial high energy strength at 3.5-5
eV, which is confirmed neither by experimental data nor by the
TDLDA results. In fact it is a consequence of a poor separable
approximation. Involving additional dipole input operators
(panel c)) weakens the artificial strength.
At the same time, the first operator is still leading.
Its contribution remains to be overwhelming in most of the RPA
states. The panel d)
demonstrates the minor role of octupole operators responsible for
the deformation coupling of the dipole and octupole modes with
$\mu =0,1$.
Our calculations performed for dipole, quadrupole
and octupole plasmons in a variety of deformed clusters
show that the deformation coupling of different multipole modes
with a given $\mu$ is usually negligible.
However, the
deformation leads to a strong splitting of the plasmons to
branches with different  $\mu$ (see panel e) and Figs. 3-6).
The comparison with experimental
data for the dipole plasmon, given in Figs. 4-5, shows that SRPA
correctly describes the deformation splitting.

Fig. 10 compares the dipole strength calculated within
exact RPA and SRPA. Cluster $Na^+_{53}$, being in focus of the
discussion above,  is chosen as a typical example for the
comparison. For the
simplicity, only $\mu =0$ dipole mode is considered. The figure
demonstrates good agreement which justifies high accuracy of
SRPA.

\vspace{0.6cm}
\noindent
{\bf Acknowledgment}\\
V.O.N. is grateful to Prof. J. Kvasil for useful discussions
of SRPA formalism. The work was partly supported (V.O.N) by
RFBR (grant 00-02-17194), Heisenberg-Landau (Germany-BLTP JINR) grant
and Votruba-Blokhintcev (Czech Republic - BLTP JINR) grant.

\newpage
\begin{center}
{\large \bf Figure captures}
\end{center}

\vspace{0.2cm}
\noindent
{\bf Figure 1:} 
Energy surface (in $Ry$) of  $Na_{119}^+$ as a function
of quadrupole and hexadecapole deformations. The iso-energy
counturs are separated by intervals 0.01 $Ry$. Three distinct
minima are marked by crosses (the energies at the minima are
given in Table I).

\vspace{0.2cm}
\noindent
{\bf Figure 2:}
Single-particle spectrum in cluster
$Na_{27}^+$ at zero and equilibrium (prolate) deformations.
In the latter case the Fermi level is ${\cal N} n_z\Lambda =321$.
The evolution of the spectrum with deformation is marked
by dashed lines.

\vspace{0.2cm}
\noindent
{\bf Figure 3:}
The same as in Fig. 2 for oblate cluster $Na_{27}^+$. The Fermi
level is ${\cal N} n_z\Lambda =310$.

\vspace{0.2cm}
\noindent
{\bf Figure 4:}
Photoabsorption cross section for dipole
plasmon in small deformed sodium clusters. The deformation
parameters are given in boxes. The experimental data
\protect\cite{exp}(triangles) are compared with SRPA results
given as bars for RPA states
(\protect\ref{eq:sig}) and as the strength function
(\protect\ref{eq:sig_av}) smoothed by the Lorentz weight.
The bars are given in $eV{\rm a}_0$.

\vspace{0.2cm}
\noindent
{\bf Figure 5:}
Photoabsorption cross section for dipole plasmon in
$Na_{53}^+$. The experimental data \protect\cite{exp}
(triangles) are compared with SRPA results for excitations
from the prolate ground state (top) and oblate isomer (bottom).
Contributions to the strength function from $\mu =0$ and 1
projections (the latter has twice larger strength) are
given by dashed curves. See Fig. 4 for notations.

\vspace{0.2cm}
\noindent
{\bf Figure 6:}
Calculated photoabsorption cross section for dipole plasmon in
$Na_{119}^+$.
The excitations from the oblate  ground state (top),
first prolate isomer (middle) and second hexadecapole
isomer (bottom) are considered. See Figs. 4 and 5
for notations.

\vspace{0.2cm}
\noindent
{\bf Figure 7:}
Calculated photoabsorption cross section for quadrupole plasmon in
prolate $Na_{27}^+$ and oblate $Na_{35}^+$. Contributions
of projections with $\mu=0,1,2$ and their sum (bottom
panel) are given as bars (\protect\ref{eq:sig})
for RPA states and as strength function
(\protect\ref{eq:sig_av}).
The bars are given in $eV{\rm a}_0$.

\vspace{0.2cm}
\noindent
{\bf Figure 8:}
The same as in Fig. 7 for octupole plasmon.

\vspace{0.2cm}
\noindent
{\bf Figure 9:}
Calculated photoabsorption cross section for
dipole plasmon in
prolate $Na_{27}^+$ and oblate $Na_{35}^+$ (see notations
in Fig. 4.). The cases include: a) unperturbed (without the residual
interaction) plasmon; b) with residual interaction
generating by leading input operator $\lambda_1 p_1=10: \quad
f_{10\mu}({\bf r})=r(Y_{1\mu}(\Omega)+Y_{1\mu}^{\dag}(\Omega))$;
c) with operators $\lambda_k p_k=10,12,14$;
d) with operators $\lambda_k p_k=10,12,14,30,32$, where two
last ones provide the coupling with octupole modes;
e) contributions of $\mu =0$ (dashed curve) and $\mu =1$
(solid curve) projections to the cross section.

\vspace{0.2cm}
\noindent
{\bf Figure 10:}
Comparison of the dipole strength with $\mu =0$
(as a squared transition matrix element (\protect\ref{eq:M}))
calculated within exact RPA (solid curve) and SRPA
(dashed curve) in prolate $Na_{53}^+$.

\newpage
\begin{table}[ht]
\caption{ Deformation parameters $\delta_2$ and $\delta_4$ (as defined in
Eq. (\protect\ref{eq:R})) and quadrupole and hexadecapole moments
$\beta_2$  and $\beta_4$ (\protect\ref{eq:QL}), calculated for the ground
and isomeric (in $Na_{53}^+$ and $Na_{119}^+$) states. For isomers
the energy deficits $\Delta E$ are  given.
}
\centering
\begin{tabular}{cccccc}\hline
 Cluster & $\delta_2$ & $\delta_4$ & $\beta_2$  & $\beta_4$ & $\Delta E$, eV \\ \hline
 $Na_{15}^+$  & 0.59  & -0.19  & 0.47  & -0.02 & - \\
 $Na_{27}^+$  & 0.33  & 0.08 & 0.36  & 0.17 & - \\
 $Na_{35}^+$  & -0.21 & 0.02 &-0.18  & 0.04& - \\
 $Na_{53}^+$  & 0.20  & -0.03  & 0.20  & $\sim 0$ & - \\
              & -0.14 & -0.09 & -0.11 & -0.06 & 0.016 \\
 $Na_{119}^+$  & -0.27 & -0.14  &-0.20  & -0.06 & - \\
                & 0.24 &-0.04 & 0.24 &  $\sim 0$ & 0.004 \\
                & -0.04 &-0.22 & -0.02 & -0.18 & 0.024 \\ \hline
\end{tabular}
\end{table}
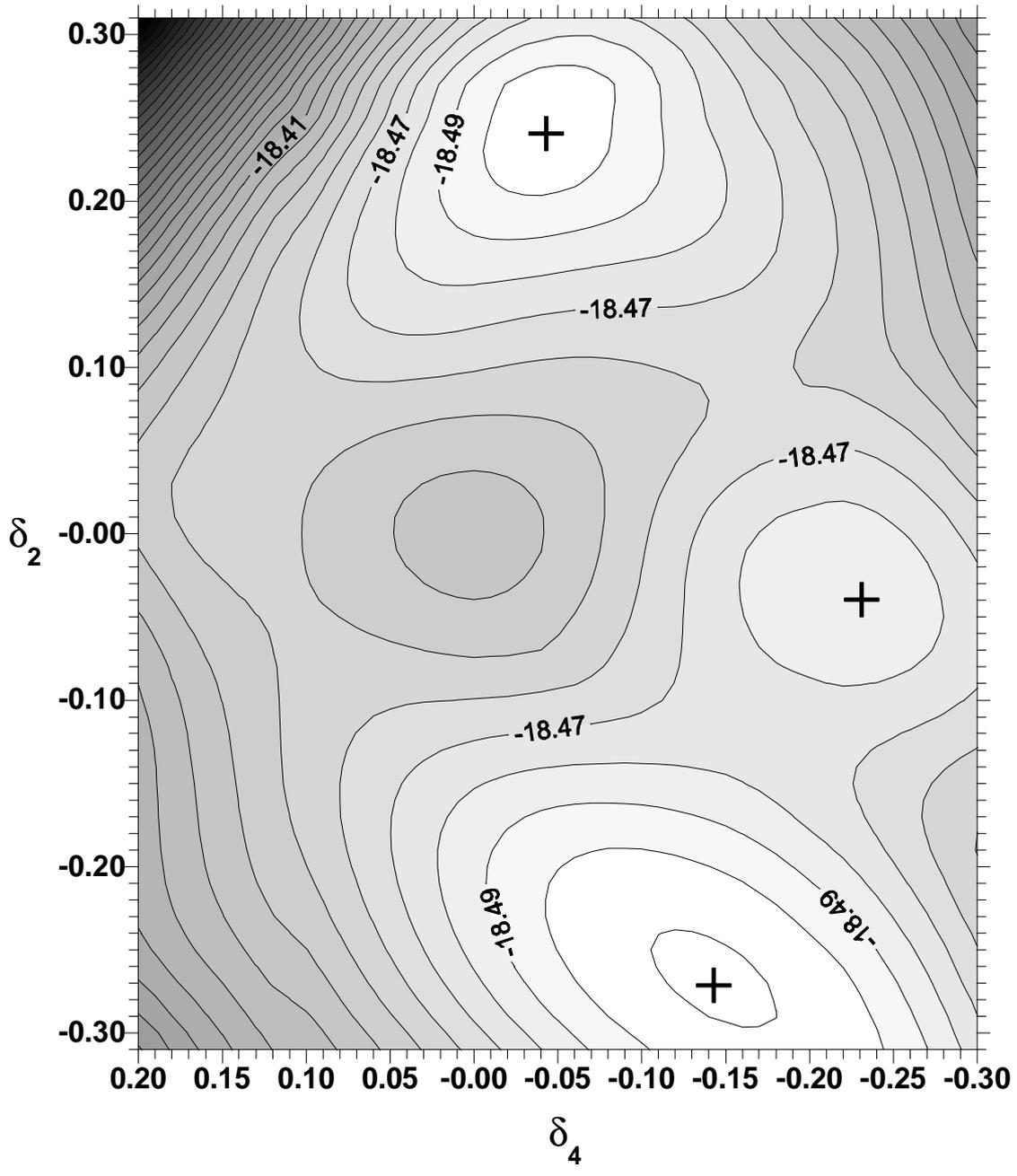
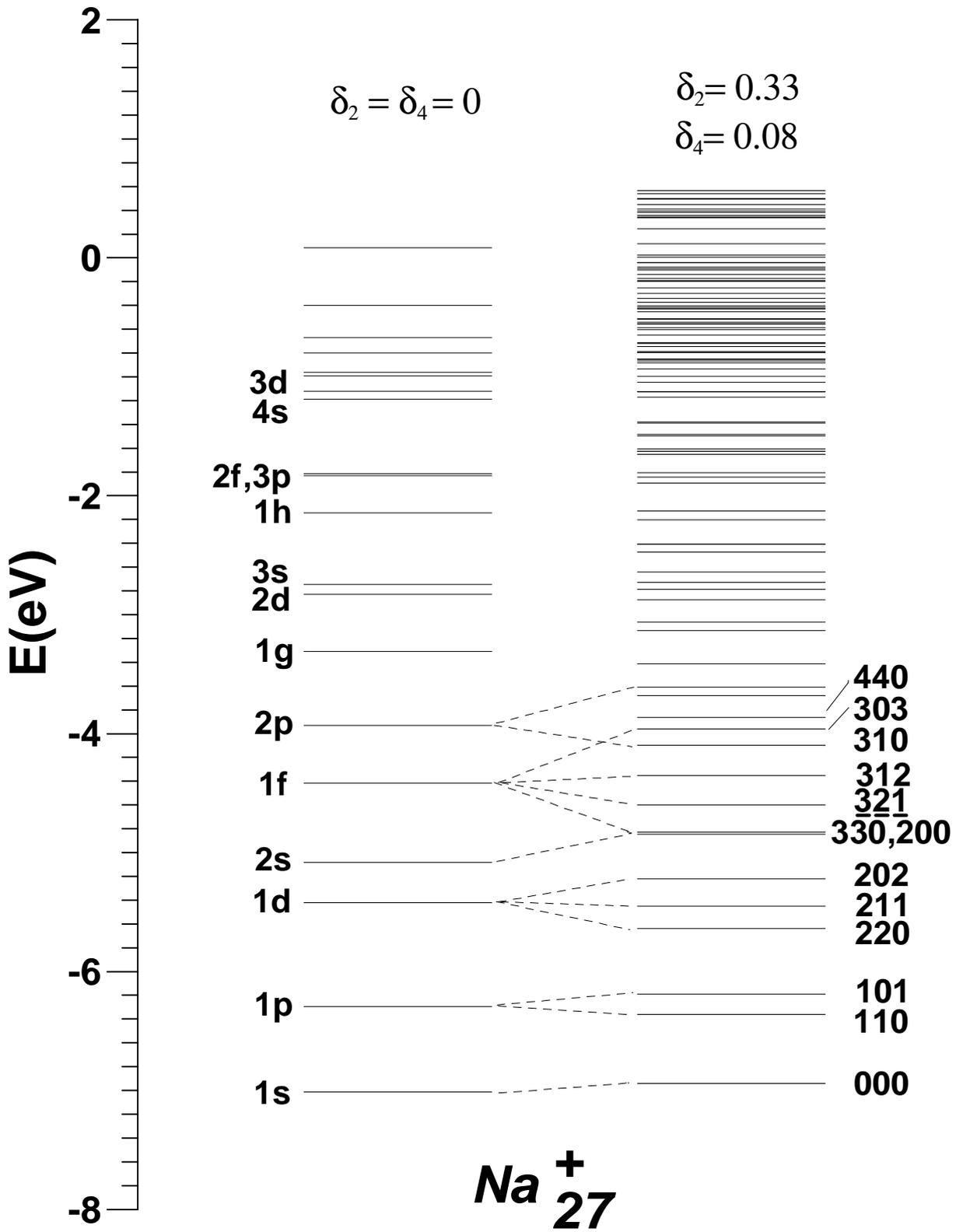
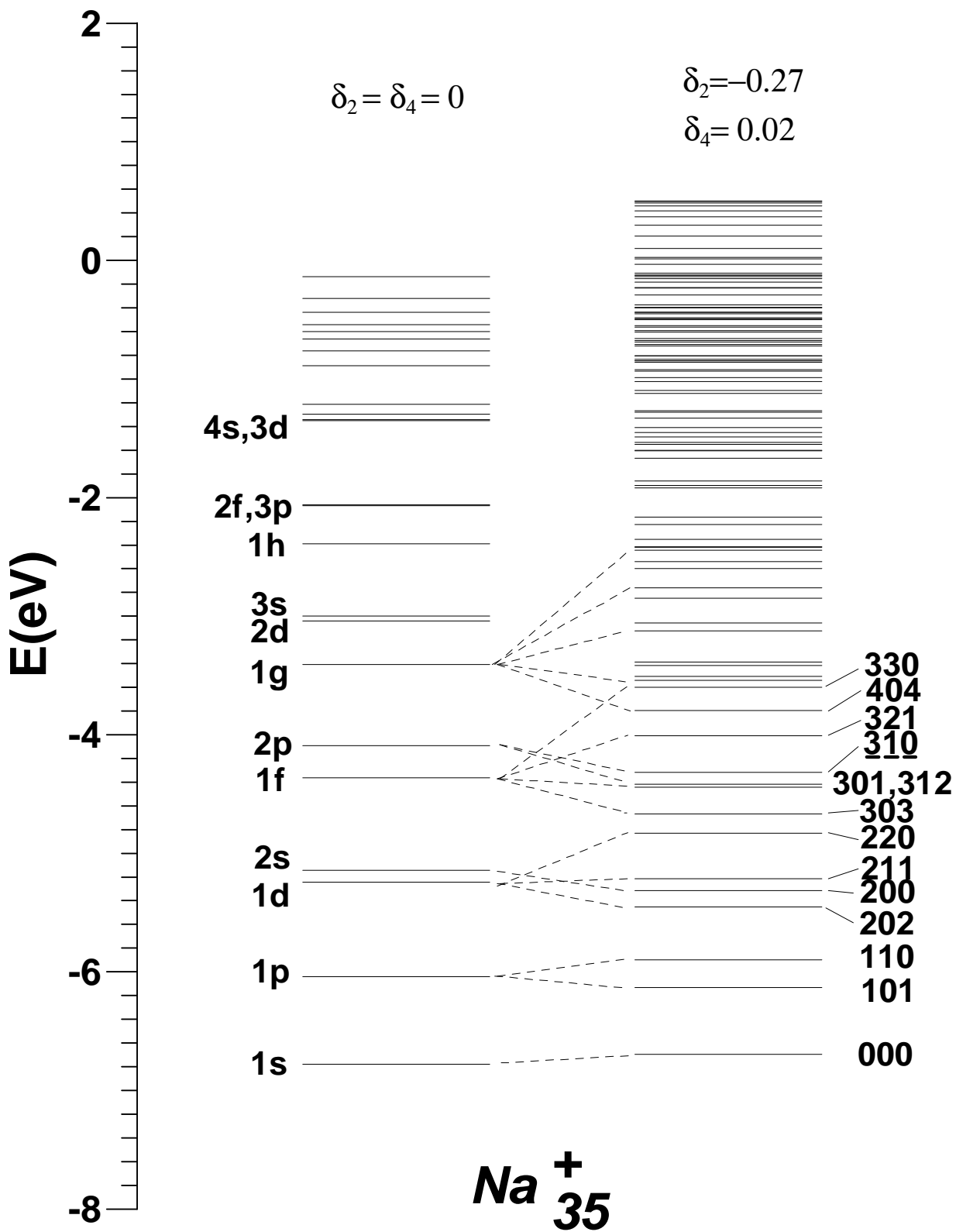
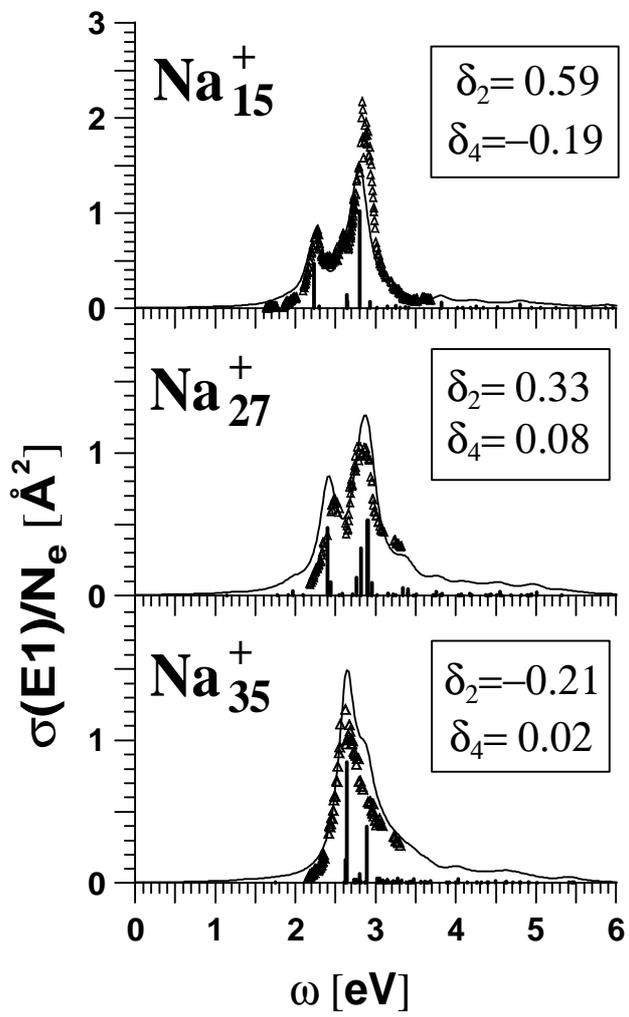
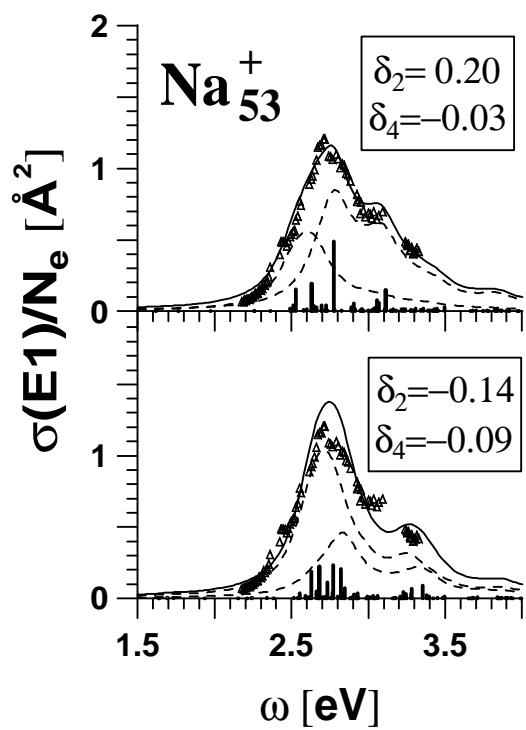
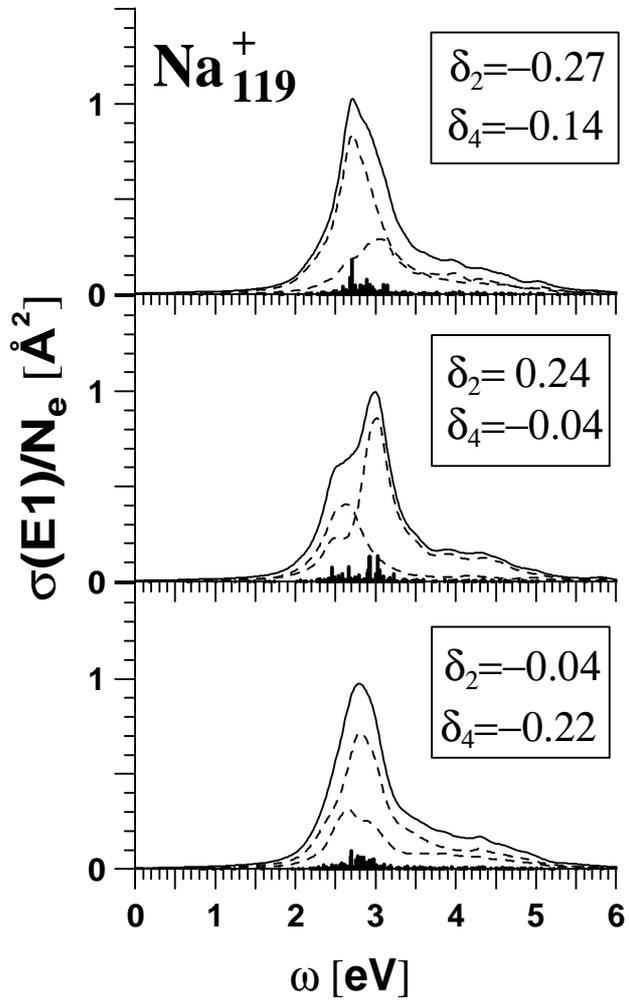
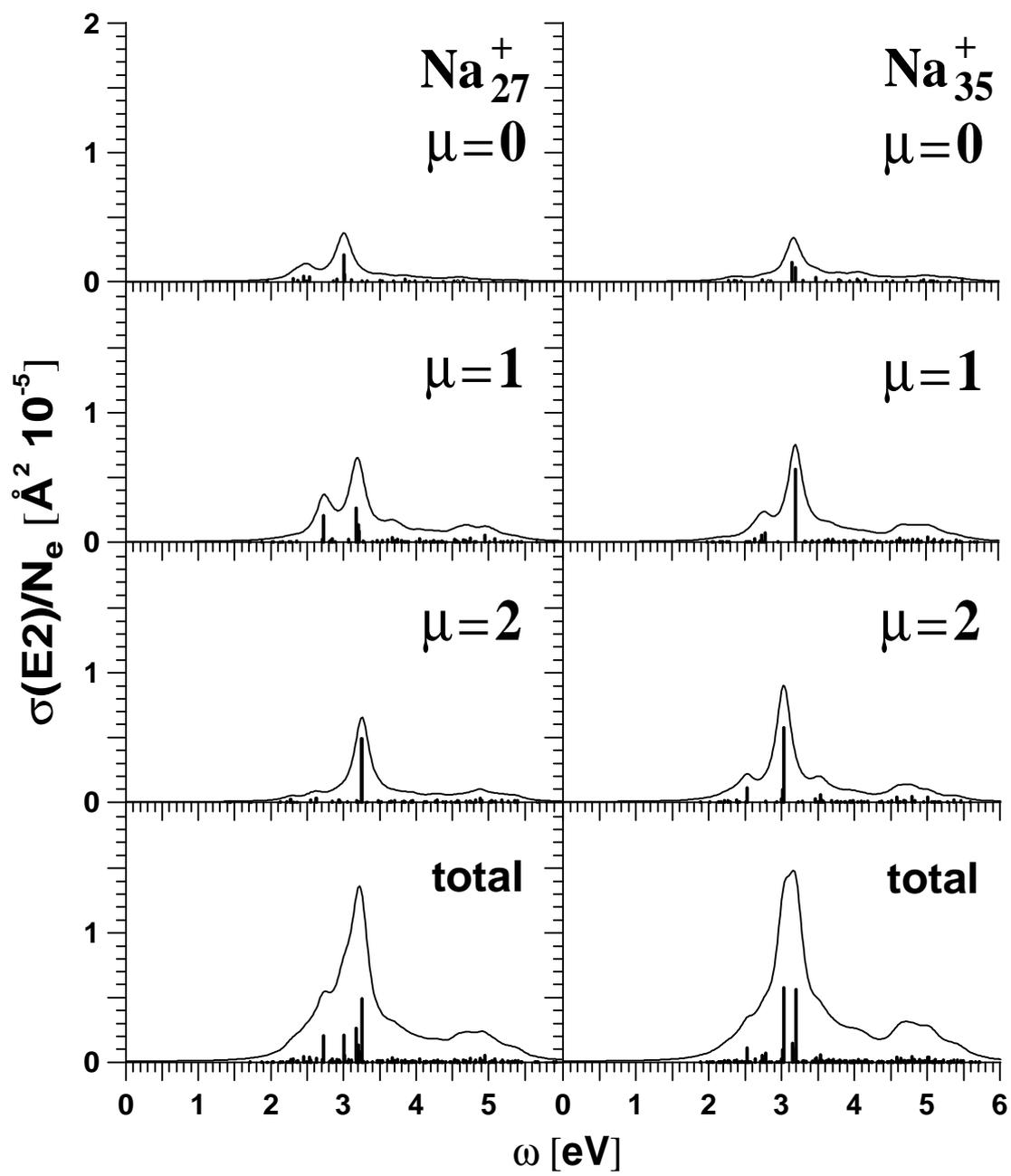
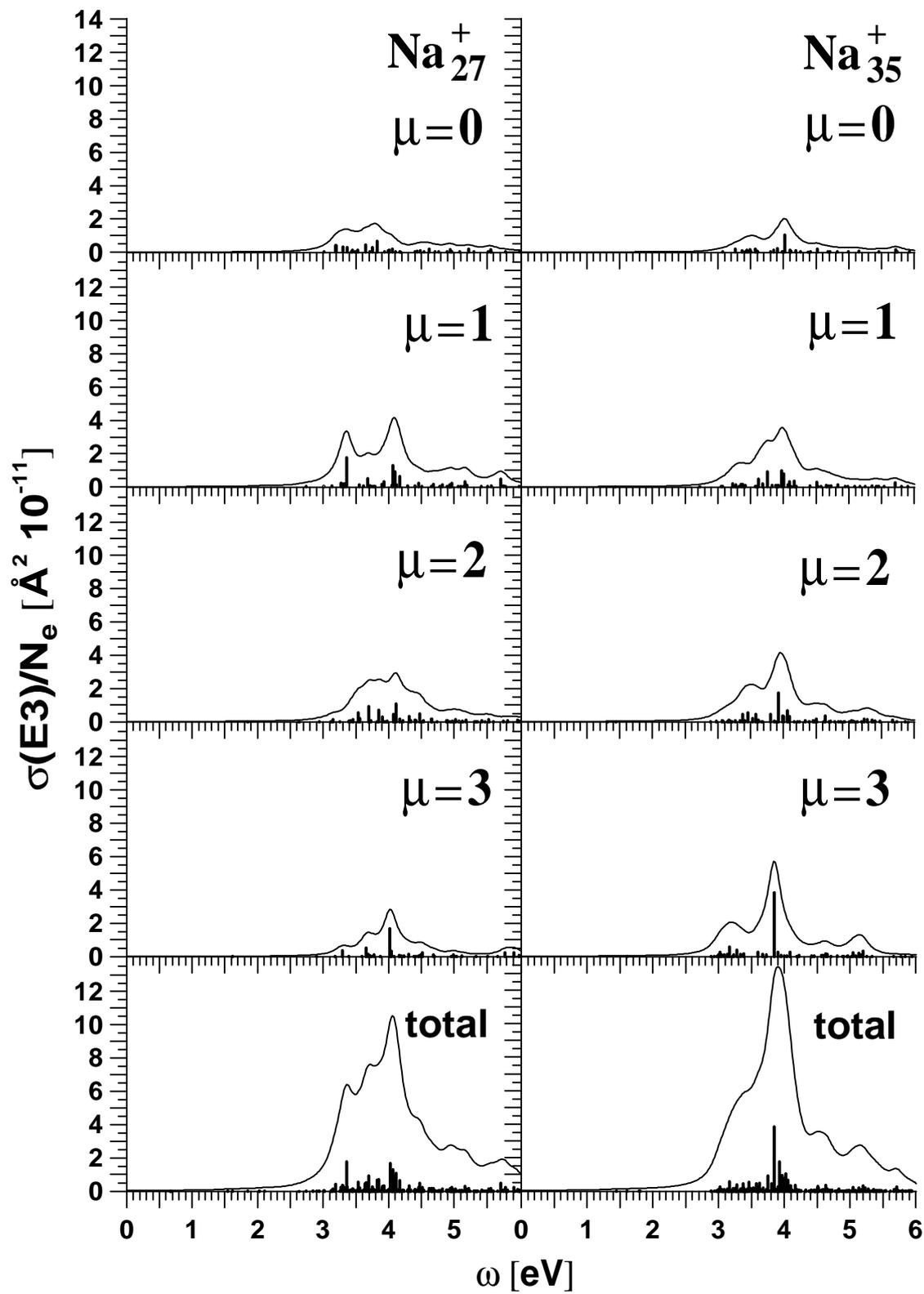
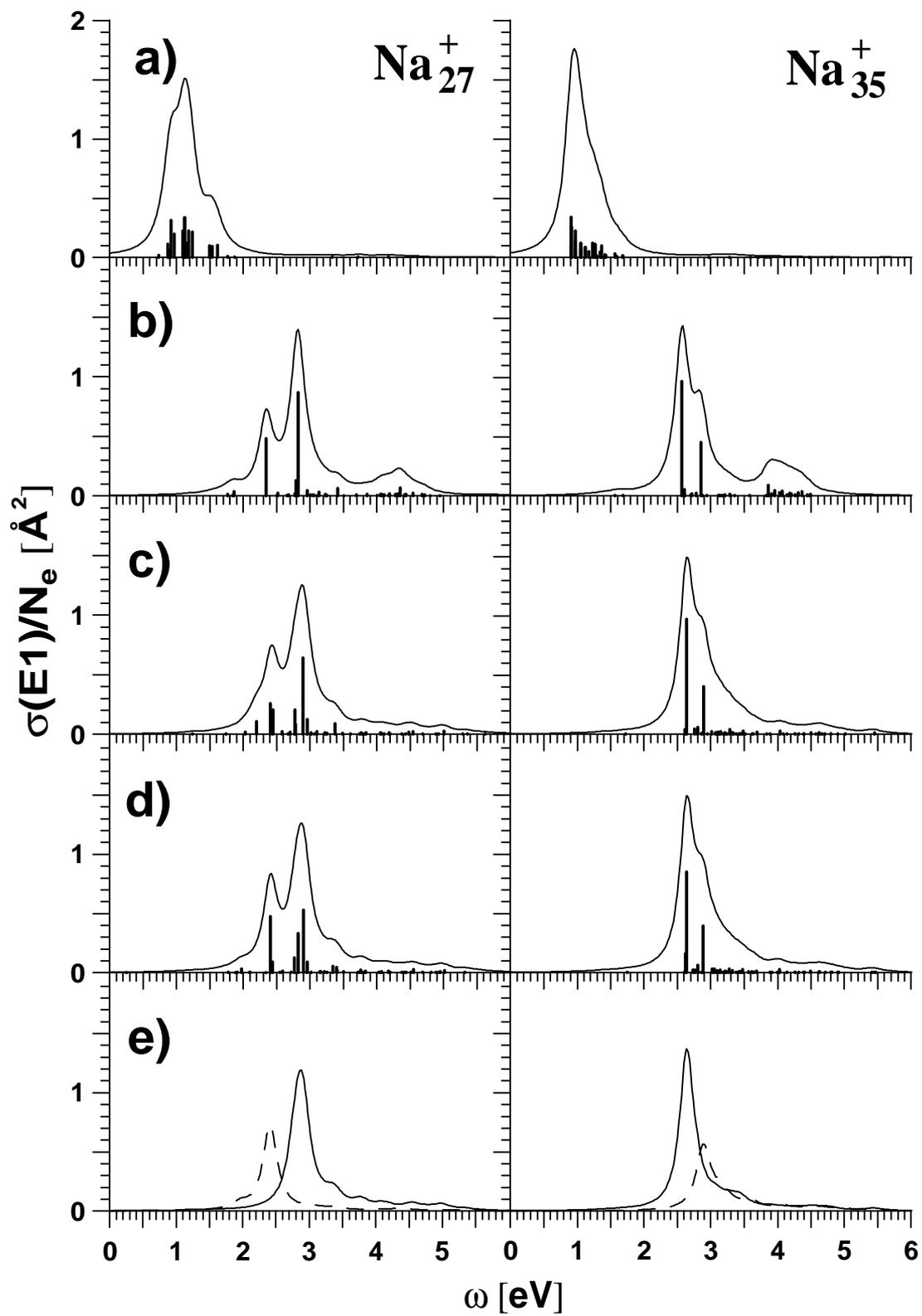
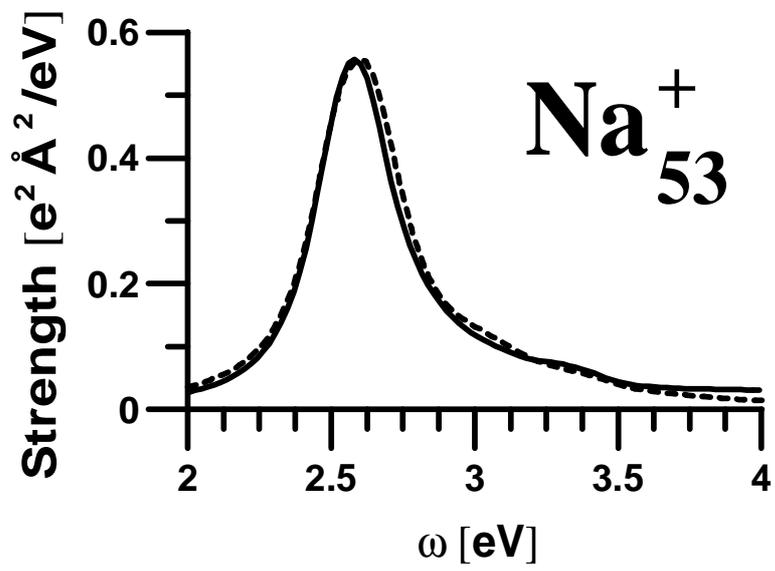
\end{document}